\documentclass{article}
\usepackage{spconf,amsmath,graphicx,hyperref}
\usepackage{booktabs}
\usepackage{tabularx}
\usepackage{xcolor}
\usepackage{float}
\usepackage{makecell} 
\usepackage{xspace}
\usepackage{multirow}
\usepackage{dblfloatfix}
\usepackage{cuted}
\usepackage{caption}
\usepackage[shortcuts]{extdash}

\title{FoleyForge: Free-form, Controllable, \\Rectified-flow Editing of Sound Effects}
\name{\begin{tabular}{c}
  Giovana Morais$^{\star,\dagger}$,
  Hugo Flores García$^\star$,
  Prem Seetharaman$^\star$,
  Johannes Imort$^\star$, \\
  Justin Salamon$^\star$,
  Magdalena Fuentes$^{\dagger}$, 
  Oriol Nieto$^\star$
\end{tabular}}
\address{
    $^\star$Adobe, $^\dagger$New York University \\ 
    \small{\textbf{\url{https://giovana-morais.github.io/2027_icassp_foleyforge_demo}}}
}
\begin{document}
\ninept

\newcommand{\sysname}{\textsc{FoleyForge}\xspace}

\maketitle


\begin{strip}
  \centering
  \vspace{-45pt}
  \includegraphics[width=\textwidth]{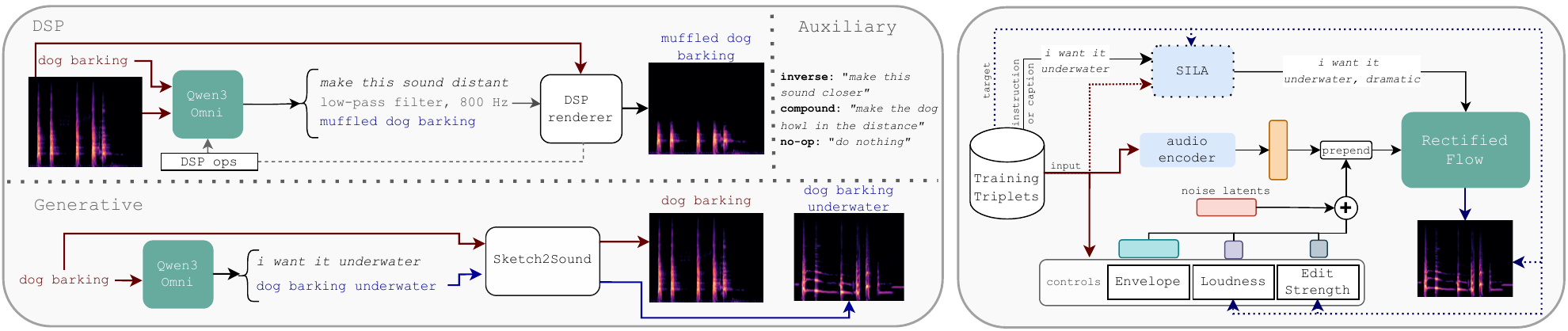}
  \vspace{-14pt}
  \captionof{figure}{\sysname overview. Left: data pipeline for DSP (top-left), generative (bottom), auxiliary (top-right) edits; input/output in red/blue, instructions in italic. Right: training and inference overview; dotted lines indicate train-only.}
  \label{fig:full_pipeline}
\end{strip}

\begin{abstract}
We present \sysname, a generative model for instruction-based editing of sound effects.
Given an input clip and a free-form instruction describing the desired change, it applies the edit while preserving the input's identity. A single model handles semantic, stylistic, and DSP-like changes.
We introduce a data pipeline that guides an audio language model to author both DSP-based acoustic edits and similarity-preserving semantic edits, thus addressing the lack of publicly available audio editing data.
Such data is then used to fine-tune a rectified-flow audio generation model, which we extend to expose attribute controls for loudness change, edit strength, and an input-derived contour envelope, each tunable via classifier-free guidance, yielding a more expressive editing model.
\sysname achieves state-of-the-art instruction adherence at competitive audio quality and source faithfulness.

\end{abstract}
\section{Introduction}
\label{sec:intro}

Text-to-audio generation now produces sound effects (SFX) of remarkable quality~\cite{evans2024stableaudioopen,kreuk2023audiogen,liu2023audioldm,liu2024audioldm2}, yet generation alone does not match how sound designers actually work. 
Sound design, especially for video content such as film, games, and short-form media, is fundamentally an \emph{iterative, per-source} process: designers layer individual sources (e.g., footsteps, impacts, whooshes, creaks) and repeatedly refine each one to lock it to the on-screen action and to creative direction: ``make the punch heavier'', ``muffle the bark'', or ``brighten the sword'' .
With current generative models, the standard way to apply such a change is to rewrite the text prompt and resynthesize from scratch. 
This discards the very identity the designer has already dialed in (e.g., its timing, timbre, and/or texture) and, crucially, its synchronization to picture, producing unpredictable, uncontrolled clips.
The image domain has already crossed this gap: instruction-based editors, whether open
\cite{brooks2023instructpix2pix,hertz2023prompttoprompt} or proprietary \cite{google_nanobanana_2026}
let a user recolor a shirt or swap a material while leaving the rest of the image
untouched. 
Similar use cases would benefit audio, yet they remain largely unexplored.


Existing approaches fall into two families. The first edits a clip toward a full \emph{target description} of the desired output~\cite{wang2023audit, manor2024zeroshotunsupervised, xu2024promptguideprecise}; the second, the strongest of which is SAO-Instruct~\cite{ungersb2025saoinstruct}, conditions directly on an \emph{edit instruction} \cite{lan2025guidingaudioeditingaudio,gao2026rfmeditingrectifiedflowmatching, liang2024wavcraft}. 
The instruction-based formulation is typically more natural
for a designer, who would rather say what to change than fully re-describe the resulting sound, and it is the setting we adopt. 
Training-free methods forgo fine-tuning: SDEdit~\cite{meng2022sdedit} noises and re-denoises the input, regenerating a new audio, while EchoEdit+~\cite{fu2026echoeditstabilizinginversionfreeaudio} is inversion-free, editing via the difference of the rectified-flow velocities under the source and target prompts. 
Both lack explicit control over edit magnitude and tend to under-edit, preserving the input at the expense of actually applying the instruction, especially when the instruction is broad. 
Beyond method design, and for more semantic/creative edits, the core obstacle is often data: supervised approaches need curated edits, and to our knowledge no such SFX dataset is publicly available.


We introduce \sysname, an instruction-driven, single-source SFX editor tailored to sound-design workflows (Fig.~\ref{fig:full_pipeline}). 
To overcome the data bottleneck, we build a synthetic data pipeline that combines complementary techniques to generate large-scale editing triplets, and use it to fine-tune a rectified-flow~\cite{liu2023flow} version of Sketch2Sound~\cite{flores2025sketch2sound}. 
The input clip is supplied through an alignment-free \emph{sequence-prepend} conditioning that imposes no fixed duration or frame-level correspondence. 
The model is trained to preserve input identity across semantic, stylistic, and acoustic edits. 
\sysname further exposes \emph{attribute controls} (e.g., loudness, edit strength) that can be steered either as continuous scalars or as natural-language magnitude phrases (e.g., ``a subtle" vs. ``a drastic" edit), following SILA~\cite{kumar2025sila} and guided by classifier-free guidance~\cite{ho2021classifier}; this yields an explicit, controllable trade-off between edit magnitude and faithfulness to the reference. 
\sysname achieves state-of-the-art instruction adherence and edit magnitude while maintaining competitive audio quality and source faithfulness, corroborated by standard automatic metrics, large audio language model judges, and a listening study.
We strongly encourage readers to listen to samples at the demo URL shown at the top of Fig.~\ref{fig:full_pipeline}.



    


\section{Method}
\label{sec:method}


\subsection{Task formulation}
\label{ssec:task}
Given an input sound effect and a free-form instruction naming the change to apply, the model produces an output that realizes the instructed change while preserving everything the instruction leaves untouched. 
The instruction describes the edit
(\emph{``make it metallic''}), not the target sound (\emph{``a metallic ball''}), and a
single model covers semantic edits, stylistic edits, and
DSP-like edits (e.g., EQ, reverb, pitch, time-stretch).
The user can also tune specific aspects of an edit (e.g., loudness) through fine-grained controls.

\subsection{Data Pipeline}
\label{ssec:data_pipeline}

Our data pipeline (Fig.~\ref{fig:full_pipeline} left) combines three types of edits: \emph{DSP} for fine-grained edits; \emph{Generative} for semantic, stylistic, and more creative ones; and \emph{Auxiliary} for robustness. 
It yields triplets of the form \emph{\{input audio, output audio, text instruction\}}.

\textbf{DSP edits} are achievable with signal processing alone.
We select DSP operations that are common tools in sound design,
such as reverb, low-pass filter, and pitch shift, and define a set of allowed
parameters, similar to~\cite{ungersb2025saoinstruct}.
We then prompt Qwen3-Omni-30B-A3B-Captioner \cite{qwen3omni}
(Qwen) to listen to the input audio and create an edit instruction achievable
\textit{only} with those operations, roughly following~\cite{chu2025text2fx}. 
The output audio is rendered by applying the chain of operations with librosa~\cite{mcfee2015librosa} and torchaudio~\cite{hwang2023torchaudio}. 

\textbf{Generative edits} cannot be achieved with DSP alone,
e.g., turning a metal door sound into a wooden door sound. 
Given sound effect descriptions from annotated data, we prompt Qwen for an edit instruction and an output caption, restricting transformations so that input and output stay within domain: action changes within the same subject, small subject changes, location changes, and material changes are allowed; unrelated substitutions are not. 
Both audios are synthesized with Sketch2Sound, either by conditioning on the input energy curve or, when the edit requires a change in energy profile, by synthesizing both from the same noise latent, which yields closely related audio.

\textbf{Auxiliary edits} are added for robustness: compound edits combine DSP and generative operations, and therefore multiple goals in a single instruction; 
inverse edits, computed both A$\to$B and B$\to$A to nudge the model toward cyclic edits (not applied in compound cases); 
and no-op edits that ask the model to reconstruct the input unchanged.

\subsection{Model}
\label{ssec:model}

\sysname's backbone (Fig.~\ref{fig:full_pipeline} right) is a rectified-flow version of Sketch2Sound, a text-to-audio model operating in the latent space of a neural audio codec, fine-tuned to allow control signals such as RMS envelope, and text-conditioned via cross-attention to FLAN-T5-XXL embeddings~\cite{chung2024scaling}.

\textbf{Source conditioning via sequence prepend.}
We condition the editor on the source clip by encoding it to \emph{clean} latent tokens (at $t{=}0$) and \emph{prepending} them as a separate, variable-length token block to the noisy output-latent sequence.
A shared positional embedding is applied over the concatenated sequence, and the transformer's self-attention lets the output tokens attend to the clean source tokens, and vice versa.
The prepended block is sliced off before the loss, so the training objective remains output-only.
Since the source enters as a variable-length \emph{prefix} rather than a channel-concatenated, time-aligned condition, the output latent is decoupled from the source's length and grid: it need not match the source's duration or be resampled or padded onto a shared latent time axis (useful, e.g., when adding reverb).
This differs from instruction-based editors such as SAO-Instruct, which concatenate the source latent to the model's input \emph{channels} and therefore register input and output on a common latent time grid, limiting the degree of the edit.

\textbf{Instruction conditioning and guidance.}
The edit instruction is encoded by the text encoder with a fixed role-tag prefix (``Edit instruction: \ldots'') and injected via cross-attention; a separate, optional output-caption slot describing the desired result enters the same way.
We use two guidance configurations: (a) the output caption is the classifier-free-guidance (CFG) axis, with the instruction always on; and (b) the \emph{instruction} is the guided axis, with the output caption retained only as a training aid.
At inference the prepended source is held in the always-on set, i.e., present in every CFG partition and never zeroed.
Thus CFG amplifies the guided text axis while the prepended source anchors input identity.

\textbf{Editing inference.}
The output latent is initialized by SDEdit~\cite{meng2022sdedit}: the encoded input is noised to an intermediate diffusion time and denoised from there, so the result stays anchored to the input's structure while realizing the instructed change.
A noise strength parameter sets how far from the input the initialization starts, so higher $t_0$ means more freedom to edit.

\subsection{Fine-grained Controls}
\label{ssec:controls}
From each \emph{\{input, output\}} training pair we derive a set of \emph{relative} controls that a user can also dial at inference. Unless noted, a control is a scalar or short vector that is projected and \emph{added to every latent frame} (a per-frame additive term).

\textbf{Loudness.} The scalar $\mathrm{LUFS}_\text{out}\!-\!\mathrm{LUFS}_\text{in}$, normalized; sets how loud the edit should be, independently of its content.

\textbf{Edit strength.} A multi-scale log-mel-spectrogram $L_1$ distance between input and output, computed on loudness-normalized audio so it is \emph{loudness-blind} and orthogonal to the loudness control. It conditions the overall degree of change.

\textbf{Energy envelope.} The median-filtered RMS curve that reflects the output's temporal amplitude contour (transients, onsets, rhythm). It is taken from the target during training and, since the target is unavailable at inference, from the \emph{input} at test time, with a guidance strength that dials how strictly this contour is followed.

\textbf{Natural-language magnitude (SILA).} We apply a SILA-style strategy~\cite{kumar2025sila} to quantize the edit strength into phrases (``a subtle / moderate / pronounced / drastic edit''), appended to the instruction based on the same log mel-spectrogram distance as in the edit strength control. This allows the amount of change to be also steerable in plain language.


  
  
  


\begin{table*}[tbh]
\centering
\resizebox{\textwidth}{!}{%
    \begin{tabular}{l ccccccc ccccccc}
    \toprule
     & \multicolumn{7}{c}{\textbf{Subtle Set} (ground truth, $n{=}100$)} & \multicolumn{7}{c}{\textbf{Creative Set} (no ground-truth, $n{=}90$)} \\
    \cmidrule(lr){2-8} \cmidrule(lr){9-15}
     & \multicolumn{5}{c}{Editing Strength} & \multicolumn{2}{c}{Audio Quality} & \multicolumn{5}{c}{Editing Strength} & \multicolumn{2}{c}{Audio Quality} \\
    \cmidrule(lr){2-6} \cmidrule(lr){7-8} \cmidrule(lr){9-13} \cmidrule(lr){14-15}
    Model & alig$\uparrow$ & faith$\uparrow$ & dir$\uparrow$ & Qwen$\uparrow$ & Gemini$\uparrow$ & FAD$\downarrow$ & PQ$\uparrow$ & alig$\uparrow$ & faith$\uparrow$ & dir$\uparrow$ & Qwen$\uparrow$ & Gemini$\uparrow$ & FAD$\downarrow$ & PQ$\uparrow$ \\
    \midrule
    \textit{Ground truth} & 0.388 & 0.747 & 0.133 & 3.28 & 4.72 & 0.340 & 6.77 & -- & -- & -- & -- & -- & -- & -- \\
    \midrule
    SAO-Instruct~\cite{ungersb2025saoinstruct} & 0.361 & 0.720 & 0.066 & 2.94 & \textbf{4.58} & 0.382 & 6.71 & 0.389 & 0.831 & 0.093 & 2.69 & 3.65 & 0.406 & 6.76 \\
    SDEdit~\cite{meng2022sdedit} & 0.293 & 0.298 & \textbf{0.127} & \underline{3.28} & 3.99 & 0.432 & 6.58 & 0.339 & 0.443 & \underline{0.094} & \underline{3.34} & \underline{4.31} & 0.423 & 6.66 \\
    EchoEdit+~\cite{fu2026echoeditstabilizinginversionfreeaudio} & \underline{0.365} & \textbf{0.902} & 0.029 & 2.52 & 4.10 & \textbf{0.349} & \underline{6.85} & 0.371 & \textbf{0.895} & 0.031 & 2.31 & 3.45 & \textbf{0.380} & \underline{6.79} \\
    \midrule
    SDEdit+Sketch (ours) & 0.339 & 0.545 & \underline{0.113} & 3.06 & \underline{4.55} & 0.387 & 6.58 & \underline{0.393} & 0.666 & 0.092 & 3.27 & 4.27 & 0.409 & 6.67 \\
    FoleyForge (ours) & 0.352 & 0.491 & 0.111 & \textbf{3.34} & 4.27 & 0.406 & 6.47 & \textbf{0.395} & 0.602 & \textbf{0.100} & \textbf{3.44} & \textbf{4.42} & 0.430 & 6.70 \\
    FoleyForge+ (ours) & \textbf{0.370} & \underline{0.856} & 0.050 & 2.36 & 4.16 & \underline{0.351} & \textbf{6.96} & 0.373 & \underline{0.889} & 0.032 & 2.47 & 3.55 & \underline{0.381} & \textbf{6.94} \\
    \bottomrule
    \end{tabular}
}
\vspace{-5pt}
\caption{Objective results for our models and baselines on Subtle and Creative sets. Best results in bold, second bests underlined.}
\label{tab:main}
\end{table*}

\section{Experiments}
\label{sec:exps}


\subsection{Objective Metrics}
\label{ssec:metrics}

Our metrics fall into two groups: how well the \emph{edit} was performed, and the
overall \emph{audio quality} of the result.

\textbf{Editing strength.}
Three metrics are cosine similarities in the LAION\-/CLAP embedding
space~\cite{wu2023laionclap}. \emph{Alignment} measures how well the output audio
matches the target caption. Because it ignores the input, a system that simply
generates a new clip matching the caption also scores well, so we read it alongside
\emph{direction}: the similarity between the audio- and text-embedding changes from
input to output. Direction rewards moving the source toward the target, penalizing
both no-op edits and off-target regenerations. \emph{Faithfulness} is the
input--output audio similarity, which is high when a system barely edits. As a purely
acoustic counterpart to these CLAP metrics, \emph{edit magnitude} (msMel,
$\uparrow$~=~larger edit) is the multi-scale log-mel-spectrogram $L_1$ distance
between loudness-normalized input and output.
Finally, \emph{adherence} measures instruction following directly using large
audio--language model (LALM) judges. Each judge sees the input clip, the edited clip,
and the instruction, and rates from 1 to 5 how well the instruction was applied while
the rest of the sound was left intact. We report the mean rating. Our primary judge is
Qwen3-Omni-Instruct~\cite{qwen3omni}. Because Qwen also wrote the edit instructions
(Section~\ref{ssec:data_pipeline}), we additionally report
Gemini~3.7~Flash~\cite{google_gemini37_2026}, which plays no other role in our
pipeline.

\textbf{Audio quality.} We report Fr\'echet Audio Distance (FAD, $\downarrow$)
between the model outputs and a fixed, proprietary reference corpus of real sound
effects. The same
reference corpus is used for both test sets, so FAD is comparable across them. For
high-level quality we report the Production Quality axis (PQ, $\uparrow$) of the
Audiobox-Aesthetics predictor~\cite{tjandra2025aes}, a per-clip, reference-free
aesthetic score averaged over the outputs.

\subsection{Setup}
\label{ssec:setup}

\textbf{Model}. We fine-tune only the rectified-flow transformer, keeping the text and audio encoders frozen, for $80\mathrm{k}$ steps on 8~GPUs with AdamW at a peak learning rate of $10^{-5}$ (300-step linear warmup, a stable phase, then linear decay over the final ${\sim}15\%$ of steps).
The text guidance signals are trained with dropout so each learns a null direction: in configuration~(a) the output caption is dropped $50\%$ of the time with the instruction always on, and in configuration~(b) the instruction is lightly dropped ($10\%$) while the output caption is kept only as a high-dropout ($50\%$) training aid.
The prepended source is likewise lightly dropped ($10\%$) during training but held always-on at inference.
Each fine-grained control is trained with dropout as well (loudness and edit strength at $0.3$, the energy envelope at $0.9$) so it learns a per-control null direction; at inference, this lets any single control axis be amplified independently via CFG while the others are held at their trained strength.


\textbf{Data}. Our final training dataset consists of 464k triplets: 173k DSP ($\sim$37\%), 118k Generative ($\sim$25\%), and 173k Auxiliary ($\sim$37\%). The Auxiliary set contains three types of edits: compound edits (59k), inverse edits (86k; not applied to compound edits); and no-op edits (28k).
The source data is an internally licensed dataset of high-quality sound-effect paired with human-annotated descriptions.
We evaluate on two benchmarks, one of more subtle edits and one of more creative ones. 
\textbf{Subtle Set} is a 100-sample, fully human-reviewed set with ground truth, split evenly between DSP and generative edits. 
It is in-distribution for our models, as it is built with the same pipeline used to generate training data, but shares no overlap with the training or validation splits. 
\textbf{Creative Set} has no ground-truth: its 90 inputs are generated by a non-finetuned Sketch2Sound and paired with edit instructions only, since the edits are too broad to be achievable through traditional DSP operations. The instructions are out-of-distribution for all models, but the audio is in-distribution since it was generated by Sketch2Sound.
All inputs and instructions were manually reviewed.

We compare \sysname against three baselines.
\textbf{SAO-Instruct}~\cite{ungersb2025saoinstruct}, the instruction-based state of the art, uses a different backbone and training data. We run its public checkpoint\footnote{\url{https://huggingface.co/disco-eth/sao-instruct}} on our benchmarks, as its test set is unreleased.
\textbf{SDEdit}~\cite{meng2022sdedit}, which noises the input and denoises it toward the instruction, and \textbf{EchoEdit+}~\cite{fu2026echoeditstabilizinginversionfreeaudio}, an inversion-free editor that differences source- and target-conditioned velocities, are \emph{training-free} methods we apply to the un-finetuned generator from which \sysname is initialized.
Because they share \sysname's architecture and differ only in \emph{how} the edit is realized, they isolate the contribution of our fine-tuning, data, and conditioning.
We also report \textbf{SDEdit+Sketch}, a training-free variant that adds the input's energy envelope to SDEdit, as in Sketch2Sound.
For \sysname, we use SDEdit noise level $t_0 = 0.9$ ($t_0{=}0$: clean, $t_0{=}1$: pure noise) and envelope guidance $\lambda_{RMS} = 0.5$, with other controls at their trained defaults ($\lambda_{text} = 3$, energy envelope grounding on); edit strength and the loudness scalar are unused.
\sysname+ uses the same checkpoint as above but applies  EchoEdit+'s training-free optimal-transport coupling, which noises the edited latent like the source so both velocity queries stay in-distribution, preventing drift along the trajectory.


\textbf{Listening study.} Twenty-five participants, screened for headphone use~\cite{woods2017headphonescreening}, rated 10 source clips under four anonymized, order-randomized conditions: SAO-Instruct, SDEdit, \sysname, and a hidden anchor presenting the unedited source as an edit.
Due to licensing issues, we randomly sample clips from the Creative Set inputs under 10\,s (to keep the test short) and limited to three systems. 
Participants heard each source alongside its edit and rated audio quality and instruction adherence on a 1--5 scale.

\section{Results and Discussion}
\label{sec:results}

\begin{figure}[]
    \centering
    \includegraphics[width=0.9\linewidth]{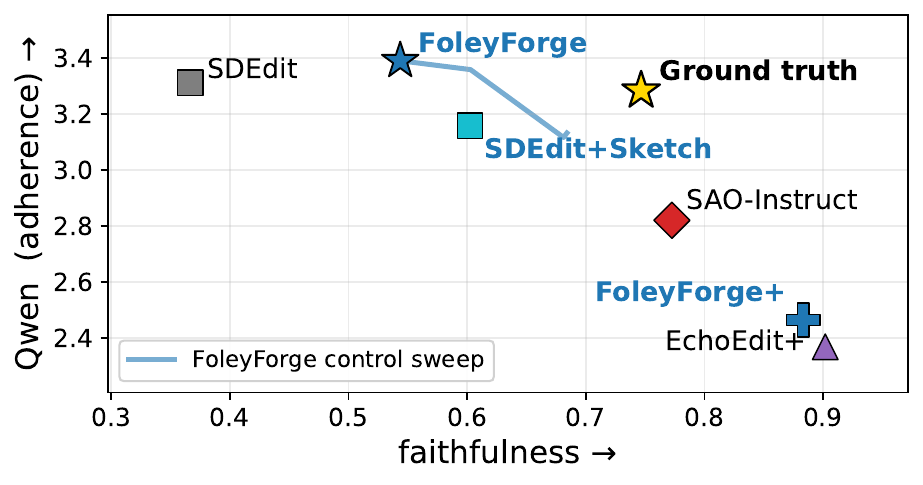}
    \vspace{-6pt}
    \caption{Adherence vs.\ faithfulness, averaged over Subtle Set and Creative Set. An ideal editor lies toward the top right.}
    \label{fig:results_combined}
\end{figure}

Fig.~\ref{fig:results_combined} plots adherence against faithfulness; an ideal editor lies toward the top right. 
The baselines each occupy a single corner: SDEdit (top left) adheres strongly, since it regenerates audio from the instruction, but nothing ties the output to the input, so faithfulness is low; EchoEdit+ (bottom right) is the most faithful but barely edits (\emph{direction} $\approx 0.03$), often producing changes too subtle to perceive; and SAO-Instruct (middle right) is likewise faithful but subtle. 
\sysname, by contrast, spans the frontier: it reaches the high-adherence end while staying more faithful than SDEdit, and sweeping the initialization noise traces the trade-off directly: higher $t_0$ raises adherence and lowers faithfulness. The same model also covers the opposite corner: \sysname+ moves into the faithful region near EchoEdit+, with the training-free SDEdit+Sketch in between at perceptible edit strengths.

\textbf{Quantitative results.}
Table~\ref{tab:main} breaks this down per test set and metric group. \sysname achieves the highest Qwen adherence on both sets and the highest \emph{alignment} on Creative Set, while \sysname+ leads \emph{alignment} on Subtle Set. \emph{Direction} splits by set: \sysname is best on Creative Set, while on Subtle Set SDEdit is nearest the reference, with SDEdit+Sketch and \sysname just behind. The two variants bracket the faithfulness axis, \sysname+ matching the most faithful baseline, EchoEdit+. Audio quality is similar across systems, so we read it as a check that editing does not degrade the output rather than as a ranking: \sysname+ has the best PQ on both sets (often above the reference edits) while \sysname sits at the low end, consistent with it moving furthest from the input.

\textbf{LALM judges.}
The signal-level metrics above measure how much the audio changed, not whether the instruction was followed, so we turn to the LALM judges for adherence. Creative Set makes the gap concrete: SAO-Instruct and \sysname have nearly identical \emph{direction} ($0.093$ vs.\ $0.100$), yet both judges separate them by a wide margin favoring  \sysname. Neither judge is a ranking on its own (SDEdit ties the reference under Qwen on Subtle Set despite a faithfulness of only $0.298$, and SAO-Instruct holds $0.831$ faithfulness on Creative Set yet still scores well below \sysname, producing edits too subtle to register) so we read the judges alongside \emph{alignment}, \emph{direction}, and faithfulness. 
On Subtle Set, \sysname is statistically indistinguishable from the reference edits under Qwen ($3.34$ vs.\ $3.28$; Wilcoxon signed-rank, $p = 0.86$), while SAO-Instruct ($p = 0.04$) and EchoEdit+ ($p < 0.001$) fall significantly below it. The judges agree on Creative Set, where \sysname leads both by a wide margin; they diverge only on Subtle Set, where Gemini places SAO-Instruct first ($4.58$ vs.\ $4.27$).

\textbf{Listening study.}
Our hidden anchor, as expected, obtains the highest audio quality ($4.54 \pm 0.80$) and the lowest adherence ($1.76 \pm 1.24$). Among the editing systems, \sysname ranks first on adherence ($3.22 \pm 1.29$), ahead of both SDEdit ($2.92 \pm 1.52$, $p = 0.011$) and SAO-Instruct ($2.69 \pm 1.41$, $p < 10^{-3}$); SDEdit and SAO-Instruct do not separate ($p = 0.109$). All tests are Wilcoxon signed-rank on per-rater, per-item pairs with Holm correction. On audio quality, \sysname places second ($3.43 \pm 1.19$), below SAO-Instruct ($3.73 \pm 1.16$, $p = 0.005$) and level with SDEdit ($3.24 \pm 1.39$). The study thus prices exactly the trade-off our design targets: \sysname trades a small but reliable amount of audio quality against SAO-Instruct for a substantially larger gain in adherence. The human ranking agrees with Qwen's on the Creating Set; licensing restrictions prevented a listening study on the Subtle Set, so the divergence there remains unresolved and is left to future work.

\begin{figure}[tb]
    \centering
    \includegraphics[width=0.9\linewidth]{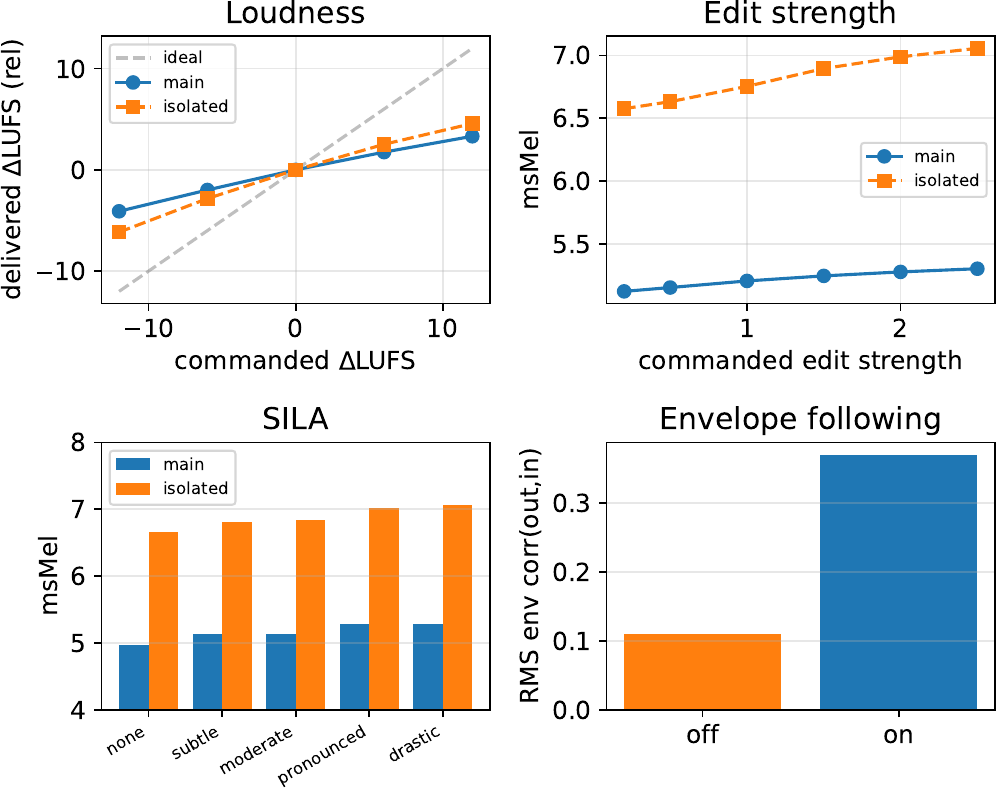}
    \vspace{-5pt}
    \caption{Control ablations. Each attribute control is monotone; the on-by-default energy envelope of the model with the main-results configuration trades control for faithfulness.}
    \label{fig:ablation}
\end{figure}

\textbf{Controllability.}
We verify that each attribute control is calibrated and monotone by sweeping one control at a time, averaged over both test sets (Fig.~\ref{fig:ablation}). We report two configurations: the \emph{main} setting from the main results (energy envelope on) and an \emph{isolated} setting that varies only the control of interest with the envelope off. \emph{Loudness} moves delivered loudness monotonically toward the commanded target in both settings, with the isolated setting tracking closer to the ideal response. \emph{Edit strength} and its \emph{SILA} phrase counterpart both raise edit magnitude (msMel) monotonically, the SILA phrases stepping cleanly from \emph{none} to \emph{drastic}. The \emph{energy envelope} grounds the output to the source: enabling it raises the input--output contour correlation ($0.11\,{\to}\,0.37$). Across all four controls, the isolated setting spans a wider range than the main setting, i.e., \emph{the always-on envelope trades control authority for source grounding}. \sysname thus exposes a controllable \emph{family} of operating points rather than a single behavior, making it flexible and easily tunable to user needs.

\section{Conclusion}
\label{sec:conclusion}
We presented \sysname, an instruction-based editor for single-source sound effects that preserves input identity across semantic, stylistic, and DSP-like edits. It combines a synthetic data pipeline, in which an LALM authors DSP-verifiable and similarity-preserving semantic edits, with a length-flexible \emph{sequence-prepend} conditioning that anchors the output to the source without tying it to the source's time grid. \sysname attains state-of-the-art instruction adherence at competitive audio quality and faithfulness. And where each baseline sits at a single fixed point on the faithfulness--adherence trade-off, \sysname's controls let the user move along it, with an inversion-free mode (\sysname+) reaching the faithful end. Rather than one more point on the curve, \sysname gives sound designers a highly flexible model tunable to their creative endeavors.

\vfill\pagebreak

\bibliographystyle{IEEEbib}
\bibliography{refs}

\end{document}